\documentclass[10pt,twocolumn]{article}

\usepackage[a4paper,margin=0.75in]{geometry}
\usepackage{graphicx}
\usepackage{amsmath,amssymb}
\usepackage{booktabs}
\usepackage{tabularx}
\usepackage{array}
\usepackage{placeins}
\usepackage{url}
\usepackage[hidelinks]{hyperref}
\usepackage{orcidlink}
\usepackage{cuted}
\usepackage{caption}
\def\acknowledgement{\par\addvspace{17pt}\small\rmfamily
\trivlist\if!\ackname!\item[]\else
\item[\hskip\labelsep
{\bfseries\ackname}]\fi}

\def\ackname{Acknowledgements}%

\newcolumntype{Y}{>{\centering\arraybackslash}X}

\begin{document}

\twocolumn[
\begin{@twocolumnfalse}

\begin{center}

{\LARGE\bfseries
WAPP: Safe Learning of Positive Security WAF Policies from Live Traffic
\par}

\vspace{0.9em}

{\large
Heba Osama\textsuperscript{1}\,\orcidlink{0000-0003-4550-9356}
\quad
Zeyad Ahmed\textsuperscript{1}
\quad
Mohamed Amgad\textsuperscript{1}
}

\vspace{0.3em}

{\large
Ahmed Saafan\textsuperscript{1}\,\orcidlink{0009-0000-7927-2292}
\quad
Jana Elfeky\textsuperscript{1}\,\orcidlink{0009-0004-8863-6333}
\quad
Mariam Abdelati\textsuperscript{2}
\quad
Haitham Ghalwash\textsuperscript{2}\,\orcidlink{0000-0001-8887-2716}
}

\vspace{0.8em}

{\small
\textsuperscript{1}Cyshield Company, Cairo, Egypt
}

\vspace{0.15em}

{\small
\textsuperscript{2}Ethical Hacking and Cybersecurity,
Coventry University -- Egypt Branch,
hosted at The Knowledge Hub Universities,
New Cairo, Egypt
}

\vspace{0.55em}

{\footnotesize
\href{mailto:heba.osama@cyshield.com}{heba.osama@cyshield.com}
\quad
\href{mailto:zeyad.ahmed@cyshield.com}{zeyad.ahmed@cyshield.com}
\quad
\href{mailto:mohamed.amgad@cyshield.com}{mohamed.amgad@cyshield.com}
}

\vspace{0.15em}

{\footnotesize
\href{mailto:ahmed.saafan@cyshield.com}{ahmed.saafan@cyshield.com}
\quad
\href{mailto:jana.elfeky@cyshield.com}{jana.elfeky@cyshield.com}
\quad
\href{mailto:mariam.abdelaati@tkh.edu.eg}{mariam.abdelaati@tkh.edu.eg}
\quad
\href{mailto:Haitham.ghalwash@tkh.edu.eg}{Haitham.ghalwash@tkh.edu.eg}
}

\end{center}

\vspace{0.8em}

\begin{abstract}
Web Application Firewalls (WAFs) mainly rely on signatures to detect known attacks, which can leave gaps against modified or previously unseen payloads. Positive security provides a complementary approach by learning legitimate traffic and blocking inputs that fall outside the learned profile. However, learning directly from live traffic can be unsafe when malicious requests contaminate the training data.
This paper presents the Whitelisting Autonomous Policy Producer (WAPP), a framework that combines trust filtering, deterministic rule synthesis, confidence scoring, and validation before enforcement. WAPP is evaluated on three controlled applications using a live Coraza and OWASP Core Rule Set (CRS) stack.
Results show that, on the tested DVWA username field, unfiltered learning becomes Degraded at 0.2\% poisoned traffic and Broken at 0.5\%, while the evaluated free text field can admit malicious inputs even without poisoning. On the frozen poisoning dataset, the ablation configuration with all seven candidate signals improves the measured poisoning resilience from 53\% to 90\%, compared with 62\% for the Kruegel--Vigna baseline. The deterministic synthesizer provides attack blocking comparable to the tested language model without model inference cost. WAPP blocks confirmed CRS bypasses on constrained fields, while free text inputs remain a precision challenge that requires character level operator control.
\end{abstract}

\vspace{0.4em}

\noindent
\textbf{Keywords:}
Web application firewall;
Positive security;
Allowlisting;
Data poisoning;
Adversarial machine learning;
Security policy generation

\vspace{1.2em}

\end{@twocolumnfalse}
]

\section{Introduction}
\label{sec:introduction}

A Web Application Firewall (WAF) inspects Hypertext Transfer Protocol (HTTP) traffic before it reaches a protected application. Traditional WAFs commonly use rules and signatures to identify known malicious patterns \cite{anuvarshini_empirical_2026}. This negative security approach is reactive because modified payloads, encoding techniques, and other evasion variants may fall outside existing rule coverage \cite{otero-mosquera_improving_2025}. Effective protection therefore requires rules to evolve as applications and attack techniques change \cite{calvo_adaptive_2022}.

Positive security reverses this logic by defining legitimate request behavior and rejecting values outside it. Statistical profiling of web request parameters can characterize properties such as value length, character distribution, token structure, and attribute presence \cite{kruegel_anomaly_2003}. A learned allowlist can use these properties to define the expected shape of individual fields without relying on recognition of a specific attack signature. For constrained inputs, this can provide strong protection against values outside the learned profile, although an overly restrictive profile may also block legitimate inputs.

The main challenge is operational. Manually creating and maintaining allowlist policies becomes difficult as applications and legitimate traffic evolve. Recent studies have demonstrated WAF detection based on machine learning using benchmark data and HTTP traffic evaluated in real time \cite{durmuskaya_web_2025,otero-mosquera_improving_2025}. However, these studies focus mainly on detecting or classifying malicious requests rather than safely learning enforceable positive security policies from live traffic. When attackers can influence the observations used for learning, data poisoning becomes part of the threat model \cite{vassilev_adversarial_2025}. This motivates evaluating both the trustworthiness of the training traffic and the behavior of the resulting policy in a live enforcement path.

This paper presents the Whitelisting Autonomous Policy Producer (WAPP), a framework that transforms live traffic into enforceable positive security WAF policies through traffic profiling, trust filtering, deterministic rule synthesis, confidence scoring, and validation before enforcement. WAPP is evaluated under adversarial conditions on three controlled applications through a live Coraza reverse proxy using the Open Worldwide Application Security Project (OWASP) Core Rule Set (CRS). The findings are limited to the applications and field types evaluated.

The work is organized around five research questions (RQs):
\begin{itemize}
    \item \textbf{RQ1 -- Learning safety:} Under what conditions does learning allowlist policies from observed live traffic become unsafe?

    \item \textbf{RQ2 -- Trust filtering:} Can trust filtering recover safe policy learning from poisoned traffic, and which signals contribute most to its effectiveness?

    \item \textbf{RQ3 -- Rule synthesis:} Given trusted traffic, does rule synthesis using a language model provide an advantage over deterministic statistical synthesis?

    \item \textbf{RQ4 -- Confidence and enforcement:} How can generated rules be scored and thresholded before enforcement, and how does this scoring relate to false positives and missed attacks?

    \item \textbf{RQ5 -- Security benefit:} Can learned positive security policies block attacks missed by WAF signature rules, and at what cost to legitimate traffic?
\end{itemize}

Together, these research questions define the scope of the study. The main contributions addressing them are summarized as follows:

\begin{enumerate}

    \item \textbf{Adversarial analysis of open population allowlist learning.}
    The analysis identifies conditions under which learning positive security policies from live traffic becomes unsafe. On the tested DVWA username field, the learner becomes Degraded at 0.2\% poisoned traffic and Broken at 0.5\%, while the evaluated free text field admits attacks even without poisoning.

    \item \textbf{Trust filtering without a clean training seed.}
    On the frozen poisoning dataset, the ablation configuration with all seven candidate signals improves measured poisoning resilience from 53\% to 90\%, compared with 62\% for the Kruegel--Vigna baseline \cite{kruegel_anomaly_2003}. Exact Shapley attribution is used to quantify the contribution of each signal.

    \item \textbf{Controlled comparison of deterministic and language model rule synthesis.}
    Under identical trusted inputs, deterministic synthesis provides attack blocking comparable to the tested language model without model inference cost and enforces closed parameter sets in six of seven tested rules.

    \item \textbf{Evidence based confidence scoring with live validation.}
    Candidate rules are scored and tested in the live WAF before enforcement. The results show that the score mainly reflects the amount of supporting evidence rather than policy safety, making representative validation necessary.

    \item \textbf{Measured additional coverage beyond signature rules.}
    Learned positive security rules block confirmed OWASP CRS bypasses on constrained fields, while experiments on rich free text identify false positives as the main limitation and evaluate character level operator control as a precision mechanism.

\end{enumerate}

The remainder of this paper is organized as follows.
Section~\ref{sec:related_work} reviews related work on WAF security, application learning, machine learning based WAFs, adversarial learning, poisoning defenses, language model based policy generation, and WAF evasion.
Section~\ref{sec:methodology} presents the WAPP methodology, including the system architecture, research design, evaluation metrics, threat model, trust filtering, rule synthesis, confidence scoring, validation, and signature bypass evaluation.
Section~\ref{sec:results} presents the experimental results corresponding to the five research questions.
Section~\ref{sec:discussion} discusses the main findings and their implications.
Finally, Section~\ref{sec:conclusion} concludes the paper and outlines directions for future work.

\section{Related Work}
\label{sec:related_work}

Prior research relevant to WAPP includes signature based WAFs, application learning, machine learning, adversarial learning, poisoning defenses, language model based policy generation, and WAF evasion.

\subsection{Signature Based and Adaptive WAFs}

Signature based WAFs, including those using OWASP CRS, detect known malicious patterns and combine rule matches through anomaly scoring \cite{owasp_crs_project_anomaly_2024,owasp_core_rule_set_project_frequently_nodate}. Although this approach provides broad protection against known attack classes, its effectiveness depends on the coverage of existing rules and transformations. Positive security follows the complementary principle of defining acceptable input and rejecting values outside the expected profile, consistent with OWASP input validation guidance \cite{owasp_cheat_sheet_series_input_nodate}.

Adaptive WAF approaches attempt to reduce the maintenance burden of static policies as applications and threat conditions change. Calvo and Beltrán \cite{calvo_adaptive_2022}, for example, use a collect analyse decide adapt loop driven by contextual risk and report fewer false positives than a restrictive static policy. Their approach adapts the protection level of an existing WAF configuration, whereas WAPP learns per endpoint positive security constraints from observed traffic and subjects generated policies to trust filtering and validation before enforcement.

\subsection{Application Learning and Web Request Anomaly Profiling}

Statistical profiling of web requests has long been used to model legitimate application behavior. Kruegel and Vigna \cite{kruegel_anomaly_2003} profile individual request parameters using characteristics such as value length, character distribution, token structure, and attribute presence to identify deviations from expected behavior. This parameter level perspective is particularly relevant to WAPP, which similarly learns field level properties and converts them into enforceable positive security constraints. The Kruegel--Vigna approach is therefore also used as an external baseline in RQ2.

Commercial WAFs have also introduced application learning and automatic policy building. FortiWeb considers successful application responses during learning \cite{fortinet_fortiweb_2023}, while F5 requires sufficient observations before learned entities are enforced \cite{f5_networks_big-ip_2018,f5_overview_2023}. Imperva incorporates factors such as source diversity, observation count, and temporal spread \cite{imperva_dynamic_2014}, and Broadcom Avi similarly requires a minimum amount of evidence before learned behavior is trusted \cite{broadcom_vmware_2024}. These approaches show that observation volume, source diversity, temporal coverage, and successful responses are meaningful indicators of learning reliability. WAPP builds on these principles but evaluates them under an adversarial setting in which the initial traffic population may already contain poisoned observations.

\subsection{Machine Learning WAFs}

Recent WAF research has explored several machine learning (ML) approaches. Supervised ML models have been used to classify malicious web requests \cite{durmuskaya_web_2025}, while feature based and hybrid learning approaches have been proposed to improve web attack detection \cite{roman_gallego_aiwaf_2025}. Otero-Mosquera et al. integrate ML with an open source WAF to improve its detection capability \cite{otero-mosquera_improving_2025}. More recently, Floris et al. proposed ModSec-AdvLearn, which combines machine learning with OWASP CRS rule selection and adversarial training to improve robustness against adversarial SQL injection attacks \cite{floris_modsec_2025}. These studies mainly focus on improving the detection or classification of malicious requests rather than generating enforceable positive security policies from observed legitimate traffic.

This differs from the objective of WAPP. A positive security policy must define what an endpoint is allowed to accept, including permitted fields, required parameters, valid types and ranges, and acceptable character patterns. WAPP therefore focuses on generating inspectable per endpoint rules from trusted traffic and validating their behavior through the live WAF before enforcement.

\subsection{Adversarial ML and Training Data Poisoning}

When policies are learned from live traffic, the training population becomes part of the attack surface. Adversarial ML research emphasizes evaluating learning systems under explicit attacker assumptions. Biggio and Roli discuss adversarial threat models for learning systems \cite{biggio_wild_2018}, while Suciu et al. characterize attacker knowledge through Features, Algorithm, Instances, and Leverage (FAIL) \cite{suciu_when_2018}. Cin\`a et al. provide a broader survey of training data poisoning attacks and defenses and organize the field around different threat models, attack strategies, and mitigation approaches \cite{cina_wild_2023}. NIST similarly organizes adversarial ML around attacker goals, capabilities, knowledge, and adaptation \cite{vassilev_adversarial_2025}. These perspectives inform the attacker model used to evaluate WAPP.

Training data poisoning is particularly relevant because malicious observations that are accepted into the learning population can influence the resulting policy. Jagielski et al. demonstrate how injected training samples can alter a learned model under explicit poisoning assumptions \cite{jagielski_manipulating_2018}. In WAPP, RQ1 characterizes when learning positive security policies from observed traffic becomes unsafe under poisoning and measures the resulting payload evasion and whitelist contamination. RQ2 then evaluates the mitigation by applying trust filtering before policy synthesis to reduce the influence of untrusted observations.

\subsection{Poisoning Defenses and No Clean Seed Learning}

A practical challenge in early policy learning is the absence of trusted clean traffic. Some poisoning defenses depend on assumptions that may not hold in this setting. For example, Jagielski et al. propose TRIM under assumptions about the poisoned fraction and the presence of a largely clean training population \cite{jagielski_manipulating_2018}. These conditions are difficult to guarantee when a WAF begins learning from its first mixed traffic window.

Cretu et al. show that anomaly sensors can sanitize their own training data without relying on a separate trusted seed \cite{cretu_casting_2008}. WAPP follows this general self sanitizing principle but uses signals available directly from WAF traffic, such as attack flags, response status, source diversity, observation volume, temporal spread, and character frequency. RQ2 evaluates these signals individually and in combination to determine whether trust filtering can improve policy learning when no clean seed is available.

\subsection{LLM Based Security Policy Generation}

Large language models (LLMs) have been explored for generating security and management policies from high level intent. Dzeparoska et al. use an LLM to decompose management intent into policy actions and validate the generated output in a controlled environment \cite{dzeparoska_llm-based_2023}. Such approaches can reduce manual policy authoring, but they also introduce additional latency, cost, and variability in the generated policy.

WAPP evaluates this trade off directly in RQ3 by comparing LLM based rule synthesis with a deterministic statistical synthesizer under the same trusted input profile and validation path. The purpose is to determine whether the additional complexity of an LLM provides a measurable advantage in the generated WAF policy.

\subsection{WAF Evasion and Security ML Evaluation}

WAF evasion remains an important limitation of signature based protection. WAF-A-MoLE demonstrates that attack preserving mutations can bypass WAF detection \cite{demetrio_waf--mole_2020}, while public PortSwigger material documents practical cross site scripting (XSS) and path traversal variants used as candidate bypasses in RQ5 \cite{portswigger_web_security_academy_cross-site_nodate,portswigger_web_security_academy_file_nodate}. WAPP does not treat signature bypass as a new problem; instead, RQ5 evaluates whether learned positive security constraints can block confirmed bypasses admitted by the tested CRS configuration.

The evaluation of security ML systems also requires realistic assumptions and appropriate baselines. Arp et al. emphasize avoiding data leakage and unrealistic experimental setups \cite{arp_dos_2022}, while TESSERACT highlights the importance of temporal realism in security evaluation \cite{pendlebury_tesseract_2019}. Sommer and Paxson further stress the need to compare learned security mechanisms with simpler alternatives and to consider operational error costs \cite{sommer_outside_2010}. These principles guide the WAPP evaluation through disjoint holdouts, live WAF replay, explicit baselines, and separate reporting of attack blocking and false positives.

\section{Methodology}
\label{sec:methodology}

\subsection{WAPP Architecture}

\begin{figure*}[!t]
\centering
\includegraphics[width=\textwidth]{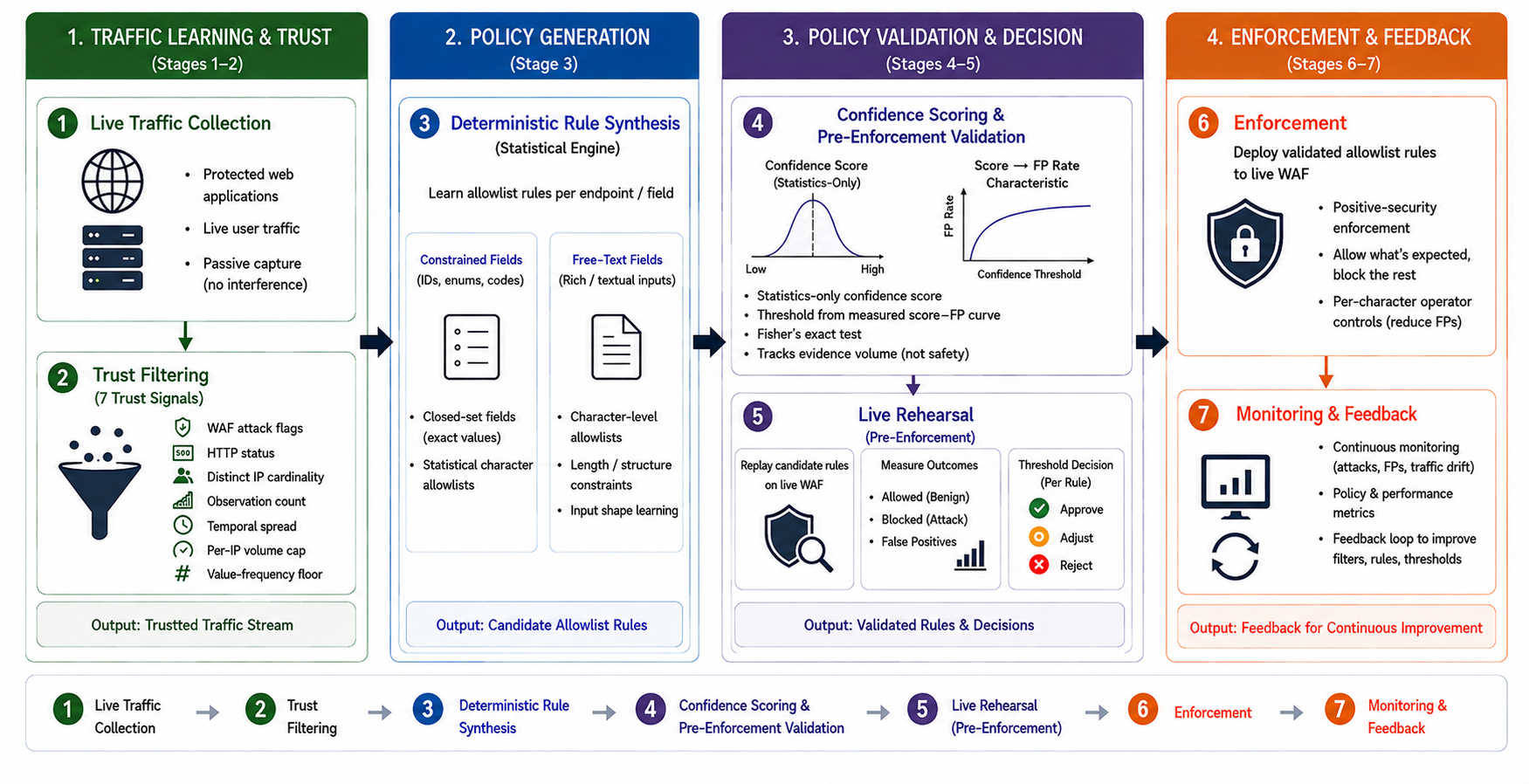}
\caption{WAPP policy learning and enforcement pipeline.}
\label{fig:pipeline}
\end{figure*}

WAPP follows a staged pipeline that separates traffic learning from policy enforcement. As shown in Figure~\ref{fig:pipeline}, the framework consists of seven stages grouped into four main phases: traffic learning and trust, policy generation, policy validation and decision, and enforcement and feedback.

Live traffic is first collected and filtered using trust signals to reduce the influence of suspicious observations. Trusted traffic is then used to generate candidate positive security rules. Before enforcement, each rule is scored, validated, and replayed through the WAF to evaluate its behavior on legitimate and adversarial traffic. Validated rules can then be enforced, while monitoring and feedback support subsequent policy adjustment and relearning.

The following subsections describe the design and evaluation of these stages in detail.

\subsection{Research Design}

The study uses an applied experimental design combining controlled traffic generation, poisoning simulation, component ablation, statistical comparison, and live WAF replay. The experiments evaluate both security behavior under controlled adversarial conditions and the behavior of generated policies in the actual enforcement path.

Two complementary evaluation modes are used:

\begin{enumerate}
    \item \textbf{Controlled analysis:} Fixed or seeded traffic populations are used to compare components under identical conditions, including poisoning experiments, trust signal ablation, and confidence calibration.

    \item \textbf{Live WAF validation:} Generated rules and legitimate and adversarial holdouts are replayed as HTTP traffic through the Coraza reverse proxy, with HTTP 403 or 406 responses counted as blocks unless otherwise stated.
\end{enumerate}

The experiments use controlled benchmark and purpose built applications rather than production traffic. Results are therefore reported for the specific applications, traffic populations, and WAF configurations evaluated.

\subsection{Metrics and Decision Criteria}

The evaluation uses four main metrics. PER is motivated by recent WAF evaluation practice that measures whether malicious payloads evade detection \cite{anuvarshini_empirical_2026}, while FPR is a standard security evaluation measure for quantifying legitimate traffic incorrectly classified as malicious \cite{hozouri_ids_2025}. WCR and PFS are defined in this study to capture contamination and combined failure in learned positive security policies.

\begin{itemize}
    \item \textbf{Payload evasion rate (PER):} the fraction of adversarial requests admitted by the learned policy. Lower values indicate better security.

    \item \textbf{Whitelist contamination rate (WCR):} the fraction of synthesized rules that accept at least one adversarial request. Lower values are better.

    \item \textbf{False positive rate (FPR):} the fraction of legitimate requests blocked by the policy. This metric is evaluated separately as an operational safety measure.

    \item \textbf{Poisoning failure score (PFS):} a metric defined in this study as the harmonic mean of PER and WCR, with a value of zero when both are zero. Higher values indicate greater failure of the learned policy under poisoning.
\end{itemize}

Poisoning resilience is also reported as
$\text{Effectiveness}=1-\text{PFS}$.

PFS is classified as \textbf{Resilient} when $\text{PFS} \le 0.10$,
\textbf{Degraded} when $0.10 < \text{PFS} \le 0.40$, and
\textbf{Broken} when $\text{PFS} > 0.40$.
FPR is evaluated independently using a threshold of $0.05$. Accordingly, a policy is considered eligible for enforcement when it satisfies the poisoning resilience criterion, remains within the FPR threshold, and passes representative pre enforcement validation.

\subsection{Data Sources and Applications}

Three applications are used in the evaluation. OWASP Juice Shop provides realistic web application traffic \cite{owasp_foundation_owasp_nodate}, while Damn Vulnerable Web Application (DVWA) provides form based traffic with session and Cross Site Request Forgery (CSRF) behavior \cite{dewhurst_damn_nodate}. Airport is a local Flask benchmark containing approximately 50 endpoints and multiple request formats.

The experimental datasets differ by research question. RQ1 uses 990 benign training records per measurable poisoning condition, with a 15 value legitimate holdout for the main live sweep and a separate 50 value holdout for the free text deconfounding analysis. RQ2 uses a deterministic Airport dataset (seed 20260601) containing 1,080 clean and 116 poisoned training records, together with 180 legitimate and 66 adversarial holdout requests. RQ3 uses 400 clean records per endpoint and 50 disjoint legitimate holdout values per endpoint. RQ4 uses a controlled synthetic value model across twelve endpoints, while RQ5 uses the live Airport application to evaluate signature bypasses.

The CSIC 2010 and ECML/PKDD datasets are included only as contextual benchmark references \cite{torrano-gimenez_http_2018,ecmlpkdd_2007_discovery_challenge_analyzing_2007}; neither is used as an experimental test set.

\subsection{Threat Model and Attacker Simulator}

The attacker model is defined across six operational dimensions:

\begin{enumerate}
    \item \textbf{Goal:} poison the learned policy, disrupt service, steal data, or probe the system \cite{biggio_wild_2018,vassilev_adversarial_2025}.

    \item \textbf{Volume and source distribution:} attacks may originate from a single high volume source or from multiple coordinated sources, including Sybil behavior \cite{fung_limitations_2020}.

    \item \textbf{Timing:} attacks may occur as short bursts, steady streams, slow traffic, or repeated waves \cite{cloudflare_what_2026}.

    \item \textbf{Payload skill:} payloads range from random fuzzing to domain aware, mimicry, boundary, and browser realistic attacks \cite{fogla_polymorphic_2006,owasp_foundation_owasp_2026}.

    \item \textbf{Knowledge:} attacker knowledge is represented using the FAIL dimensions, with none, partial, or full knowledge of each component \cite{suciu_when_2018,vassilev_adversarial_2025}.

    \item \textbf{Adaptation:} attackers may be static, react after observing system behavior, or adapt continuously during the attack \cite{biggio_wild_2018,vassilev_adversarial_2025,ennaji_adaptive_2026}.
\end{enumerate}

The attacker simulator is deterministic, allowing the same seed and configuration to reproduce the same generated records. It is used for poisoned traffic generation in RQ1 and RQ2, while RQ4 reuses selected seeded payload generators. RQ3 and RQ5 use separate live replay paths. This separation avoids assuming that all research questions use the same dataset or attack execution.

For RQ1, poisoned records are inserted as successful, unflagged WAF observations. This models the case in which malicious traffic has already passed the existing WAF and is therefore eligible for policy learning. Requests already blocked by the WAF are not included because they cannot poison the learned baseline. The RQ1 open population baseline disables the trust filtering mechanisms, including the 0.90 value frequency floor, so that the experiment measures learning directly from the observed request population.

\subsection{Trust Filtering}
\label{sec:trust}

Before rule synthesis, WAPP evaluates seven configurable trust signals: WAF attack flags with a per source strike ban, successful HTTP status, distinct source IP cardinality, total observation count, temporal spread, an optional per IP volume cap, and a per parameter value frequency floor. All signals operate only on the current training window and do not require a trusted clean seed, historical tenant baseline, or tenant specific reputation feed. This first window setting is the focus of RQ2.

Each signal is controlled independently by a runtime flag and a strictness parameter. The default configuration enables the first five signals and the value frequency floor, giving six active signals, while the per IP volume cap is disabled by default and uses a threshold of 30 when enabled. For evaluation, the ablation analysis tests all $2^7 = 128$ possible signal combinations on the same frozen dataset. The reported results include the empty filter, seven individual signals, all $21$ pairs, and the configuration with all seven signals enabled.

Ground truth labels are used only for evaluation and are not available to the trust filter or rule synthesizer \cite{cretu_casting_2008}. Filter level false positive and false negative rates are evaluated separately from policy contamination, PER, legitimate FPR, and PFS because endpoint level gating and character level hardening can improve poisoning resilience without necessarily removing individual poisoned records. The seven trust signals, their default settings, grounding, and roles in WAPP are summarized in Table~\ref{tab:wapp-signals}.

\begin{table*}[!htt]
\centering
\caption{WAPP Signals, Defaults, and Grounding}
\label{tab:wapp-signals}
\small
\setlength{\tabcolsep}{6pt}
\begin{tabularx}{\textwidth}{
@{\hspace{0.03\textwidth}}
  >{\raggedright\arraybackslash}p{115pt}
  >{\centering\arraybackslash}p{75pt}
  >{\raggedright\arraybackslash}X
  >{\raggedright\arraybackslash}X
}
\toprule
\textbf{Signal} & \textbf{Default} & \textbf{Grounded in} & \textbf{Role in WAPP} \\
\midrule
WAF attack flags + per source ban & On / 3 strikes & CRS anomaly scoring \cite{owasp_crs_project_anomaly_2024}; IP reputation \cite{cloudflare_block_2026}. & Drop flagged records and suppress repeatedly flagged sources. \\
\addlinespace
HTTP status allowlist & On / success & Successful response learning \cite{fortinet_fortiweb_2023}, \cite{f5_networks_big-ip_2018}, \cite{imperva_dynamic_2014}. & Learn only from successful application responses. \\
\addlinespace
Distinct IP cardinality & On / {\normalfont$\ge$}50 IPs & Imperva default 50; F5 learning gate \cite{f5_networks_big-ip_2018}, \cite{imperva_dynamic_2014}. & Require cross source evidence before synthesis. \\
\addlinespace
Observation count & On / {\normalfont$\ge$}50 obs. & Imperva 50; Avi 100; FortiWeb 400 \cite{fortinet_fortiweb_2023}, \cite{imperva_dynamic_2014}, \cite{broadcom_vmware_2024}. & Require sufficient repeated evidence. \\
\addlinespace
Temporal spread & On / {\normalfont$\ge$}12 h & Imperva 12 h; temporal evaluation discipline \cite{pendlebury_tesseract_2019}, \cite{imperva_dynamic_2014}. & Avoid learning short bursts or scans as stable behaviour. \\
\addlinespace
Per IP volume cap & Off / 30 if enabled & Rate limiting and RFC 6585 \cite{owasp_foundation_owasp_2026}, \cite{cloudflare_rate_2026}, \cite{internet_engineering_task_force_rfc_2012}. & Optional limit on dominance by one source. \\
\addlinespace
Value frequency floor & On / 0.90 & Positive security profiling \cite{kruegel_anomaly_2003}, \cite{f5_networks_big-ip_2018}, \cite{imperva_dynamic_2014}. & Admit a special character only when it appears in {\normalfont$\ge$}90\% of observed values; poison remains in the tally. \\
\bottomrule
\end{tabularx}
\end{table*}

\subsection{Trust Filter Action Levels}

The seven trust signals operate at different levels, so the filter level False Negative Rate (FNR) alone does not capture their full effect. Their actions fall into three categories:

\begin{itemize}
    \item \textbf{Record level filtering:} WAF attack flags, HTTP status, temporal spread, and the optional per IP volume cap can remove individual observations before synthesis.

    \item \textbf{Endpoint level gating:} distinct IP cardinality and total observation count can prevent rule synthesis when an endpoint has insufficient supporting evidence, even if no individual record is removed.

    \item \textbf{Character level hardening:} the value frequency floor does not remove records but restricts the character set admitted by the generated rule.
\end{itemize}

As a result, a signal may have a record level FNR of 1.00 and still reduce whitelist contamination. Endpoint gates can prevent an unsafe rule from being generated, while character level hardening can restrict the resulting policy even when poisoned records remain in the training set.

\subsection{Rule Synthesis}

Rule synthesis converts the trust filtered profile into one enforceable policy per endpoint. Records are grouped by endpoint and method, dynamic path segments are normalized, and request bodies are parsed across JSON, form, multipart, XML/SOAP, GraphQL, and plain text. Per field constraints are then derived for type, requiredness, range, character evidence, and parameter sets.

The deterministic engine applies typed patterns where recognized and otherwise uses the per field value frequency floor, set to 0.90 by default, to control admitted special characters. This default applies to the WAPP synthesis configuration used in RQ2 onward; the open population baseline in RQ1 uses the same deterministic synthesizer with the value frequency floor disabled. For comparison, the LLM receives the same trusted profile and produces the same rule schema using a fixed qwen/qwen3.5-27b configuration through OpenRouter. The LLM arm is repeated five times per application, while the deterministic arm runs once. Both use the same conversion, enforcement, attack replay, and legitimate holdout paths, making the synthesis method the main experimental variable. The deterministic approach is grounded in per parameter web anomaly modeling \cite{kruegel_anomaly_2003}, while the LLM comparison follows prior policy generation work \cite{dzeparoska_llm-based_2023}.

Before accepting an RQ3 result, the comparison is checked for disjoint training and holdout data, valid generated character classes, correct replay counts, clean rule state between runs, and compliance with the LLM cost budget. These checks reduce the risk of confounding from data leakage, malformed rules, stale configurations, or incomplete replay.

\subsection{Validation, Scoring, and Enforcement}

Candidate rules receive an evidence score in $[0,1]$ based on distinct source IPs, observation count, active hours, per field thinness, and minimum evidence requirements. The score is advisory and does not automatically authorize enforcement. Each candidate rule is compiled into the live WAF in shadow mode and replayed against a disjoint legitimate holdout and a fixed attack suite. False positives or missed attacks can reduce the score through configurable feedback penalties.

The main calibration uses 108 synthetic legitimate values and 13 genuine attack vectors per endpoint. A wider 17 item suite additionally includes four non exploit boundary and mimicry probes. Because the synthetic holdout has a limited character range, the DVWA login rule is also tested using a representative holdout containing separator characters before enforcement conclusions are made.

This evidence based validation is consistent with automatic policy building practice \cite{f5_overview_2023}, while the use of disjoint live replay follows out of sample security evaluation guidance \cite{arp_dos_2022,sommer_outside_2010}.

\subsection{Signature Bypass Evaluation}

The signature bypass experiment uses Coraza with OWASP CRS 4.25.0 at paranoia level 1 on the Airport product search route. The local anomaly threshold is 100, while the CRS recommended threshold of 5 is also tested. Canonical SQL injection, cross site scripting (XSS), and path traversal payloads first confirm that CRS is actively blocking attacks at threshold 5. Eleven bypass candidates from the same attack classes are then replayed, and those admitted by CRS form the confirmed bypass set.

To isolate the effect of the learned positive security policy, confirmed bypasses are replayed at threshold 100, where the signature layer does not block them, and again at threshold 5 to evaluate the layered configuration. The five confirmed bypasses are replayed against two constrained fields, producing ten attacker requests. Two in shape benign values, one for each field, are used as negative controls. CRS anomaly scoring is documented by the CRS project \cite{owasp_crs_project_anomaly_2024,owasp_core_rule_set_project_frequently_nodate}, while the XSS and path traversal candidates are based on PortSwigger references \cite{portswigger_web_security_academy_cross-site_nodate,portswigger_web_security_academy_file_nodate}.

\section{Results}
\label{sec:results}

\subsection{Open Population Learning Is Unsafe Without Filtering}

Before the main three application evaluation, RQ1 validates the metrics on a seeded login username field using 800 clean training records, a 200 value legitimate holdout, and ten attack payloads. Poison is defined as a fraction of the final poisoned training set. The pilot shows that PER and PFS rise sharply when poison is introduced, while FPR remains 0, demonstrating that poisoning can compromise the learned policy without causing visible false positive failures. Full pilot results are reported in Table~\ref{tab:pilot}.
\begin{table}[!ht]
\centering
\caption{RQ1 metric validation pilot}
\label{tab:pilot}
\scriptsize
\setlength{\tabcolsep}{2.5pt}

\begin{tabular}{@{}ccccccc@{}}
\toprule
\textbf{Poison \%} &
\textbf{Injected} &
\textbf{PER} &
\textbf{WCR} &
\textbf{PFS} &
\textbf{Tier} &
\textbf{FPR} \\
\midrule

0.00  & 0   & 0.00 & 0.00 & 0.000 & Resilient & 0.00 \\
0.99  & 8   & 0.90 & 1.00 & 0.947 & Broken    & 0.00 \\
2.91  & 24  & 1.00 & 1.00 & 1.000 & Broken    & 0.00 \\
4.99  & 42  & 1.00 & 1.00 & 1.000 & Broken    & 0.00 \\
9.91  & 88  & 1.00 & 1.00 & 1.000 & Broken    & 0.00 \\
20.00 & 200 & 1.00 & 1.00 & 1.000 & Broken    & 0.00 \\

\bottomrule
\end{tabular}
\end{table}
The main RQ1 evaluation shows that the tested DVWA login username field is Resilient under clean training but becomes unsafe once poisoned traffic enters the learning population. At 0\% poisoning, the learned policy blocks all ten attack payloads (PER 0.00 and PFS 0.00). At 1\%, 3\%, 5\%, and 10\% poisoning, PER and PFS reach 1.00. A finer three seed sweep places the transition below 1\%: at 0.2\% poisoning, mean PER is 0.1667 (SD 0.0577), classified as Degraded, while at 0.5\% it reaches 0.50 (SD 0.10), classified as Broken.

The Airport free text comment field is already Broken without poisoning, with PER 0.30 and PFS 0.462 because legitimate and malicious inputs share punctuation. Its 15 value live holdout gives FPR 0.00, whereas a separate 50 value deconfounding holdout gives FPR 0.26. The Juice Shop feedback field is excluded from this poisoning sweep because unresolved captcha requests return HTTP 500 before a WAF verdict. It is evaluated later in RQ3 after the captcha requirement is satisfied. Table~\ref{tab:open-population-sensitivity} summarizes these results.

\begin{table*}[!ht]
\centering
\caption{RQ1 open population poison sweep outcome by field}
\label{tab:open-population-sensitivity}
\small
\begin{tabularx}{\textwidth}{@{} 
>{\raggedright\arraybackslash}p{85pt}
>{\raggedright\arraybackslash}X
>{\raggedright\arraybackslash}X
>{\raggedright\arraybackslash}X
>{\raggedright\arraybackslash}X @{}}
\toprule
\textbf{App / field} &
\textbf{0\% result} &
\textbf{Sub 1\% bracket} &
$\geq$\textbf{1\% result} &
\textbf{Legitimate FPR / notes} \\
\midrule

Airport free text comment &
PER 0.30; PFS 0.462; \textbf{Broken} &
Already Broken at 0\% &
PER 1.00; PFS 1.00; \textbf{Broken} &
0.00 on 15 value live holdout; 0.26 (13/50) on a separate 50 value deconfounding holdout (PER 0.20 in that arm) \\

\midrule

DVWA login username &
PER 0.00; PFS 0.00; \textbf{Resilient} &
0.2\%: mean PER 0.1667, SD 0.0577, \textbf{Degraded}; 
0.5\%: mean PER 0.50, SD 0.10, \textbf{Broken} (3 seeds) &
PER 1.00; PFS 1.00; \textbf{Broken} &
0.00 on 15 value live holdout \\

\midrule

Juice Shop feedback &
Excluded because captcha returns pre WAF HTTP 500 &
Not measured &
Excluded &
Measured later in RQ3 with the captcha satisfied \\

\bottomrule
\end{tabularx}
\end{table*}

The poison sweep was then repeated using the composed WAPP defense: the RQ2 trust filter at its default configuration, the locked statistical synthesizer, and RQ4 advisory scoring. The composed defense recovered poisoning resilience on the measured cells, with PER remaining 0.00 across poisoning levels from 0\% to 10\%. However, the 0.90 value frequency floor caused substantial false positives. FPR reached 1.00 for the Airport free text comment field and 0.80 (12/15) for the DVWA login holdout. Both cells were therefore classified as Resilient by PFS while still violating the 0.05 FPR guardrail.

These results show that recovering poisoning resilience does not by itself guarantee an operationally usable policy. The poisoning metric and the false positive guardrail must therefore be considered together.

\newpage 

\subsection{The Stacked Trust Filter Sharply Reduces Contamination}

RQ2 evaluates the trust filter on a frozen synthetic dataset containing 1,080 clean and 116 poisoned training records, with separate legitimate and adversarial holdouts. With no filtering, 8 of 11 synthesized rules are contaminated, giving WCR 0.7273, PER 0.3485, and effectiveness 0.5288. With all seven candidate signals enabled for the ablation experiment, WCR decreases to 0.1111 and PER to 0.0909, while effectiveness increases to 0.9000. Legitimate FPR remains 0.0222 on the structured 180 request holdout. The Kruegel--Vigna baseline \cite{kruegel_anomaly_2003} improves over the empty filter but reaches a lower effectiveness of 0.6231, with WCR 0.5455 and PER 0.2879.

Figure~\ref{fig:rq2-config-comparison} summarizes the comparison. Although 40 of 116 poisoned records survive the record level filter ($\text{FNR}_{\text{filter}}=0.3448$), endpoint gating and character hardening further reduce policy contamination. The structured holdout does not contain punctuation rich free text, so its FPR of 0.0222 should not be generalized to free text fields.

\begin{strip}
\centering
\includegraphics[width=0.82\textwidth]{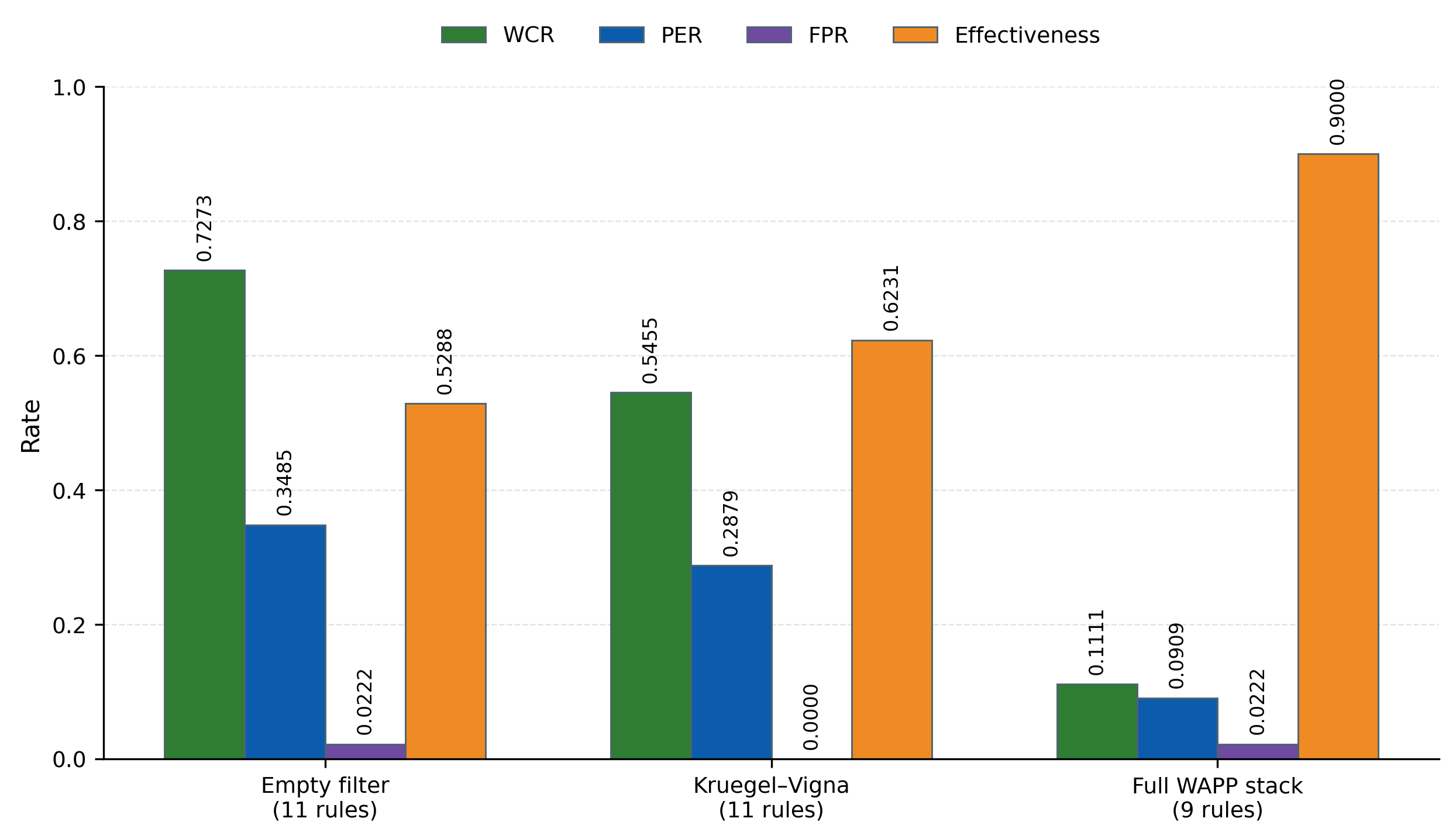}

\captionof{figure}{RQ2 filtering configuration comparison.}
\label{fig:rq2-config-comparison}
\end{strip}

The full $2^7=128$ coalition analysis shows that the value frequency floor has the largest Shapley contribution (0.1494), followed by WAF attack flags and HTTP status (0.0968 each). The per IP cap contributes 0 across the tested profiles \cite{shapley1953value}. Table~\ref{tab:trust-signals-ablation} summarizes the standalone results.

\begin{table*}[!ht]
\centering
\caption{RQ2 signal ablation results}
\label{tab:trust-signals-ablation}
\small
\setlength{\tabcolsep}{1.5pt}
\renewcommand{\arraystretch}{1.2}

\begin{tabularx}{\textwidth}{
  >{\raggedright\arraybackslash}p{52pt}
  >{\centering\arraybackslash}X
  >{\centering\arraybackslash}X
  >{\centering\arraybackslash}X
  >{\centering\arraybackslash}X
  >{\centering\arraybackslash}X
  >{\raggedright\arraybackslash}p{55pt}
}
\toprule
\textbf{Configuration} &
\textbf{WCR} &
\textbf{PER} &
$\mathbf{FNR}_{\mathbf{filter}}$ &
\textbf{Effect.} &
\textbf{Shapley} &
\textbf{Main effect} \\
\midrule

Empty filter &
0.7273 & 0.3485 & 1.0000 & 0.5288 & n/a &
none \\

\addlinespace

Per IP cap (disabled by default) &
0.7273 & 0.3485 & 1.0000 & 0.5288 & 0.0000 &
no contribution at tested poison level \\

\addlinespace

Distinct IP cardinality &
0.6667 & 0.3182 & 1.0000 & 0.5692 & 0.0094 &
low evidence / invented endpoints \\

\addlinespace

Observation count &
0.6667 & 0.3182 & 1.0000 & 0.5692 & 0.0094 &
low evidence / invented endpoints \\

\addlinespace

Temporal spread &
0.6667 & 0.3182 & 0.6552 & 0.5692 & 0.0094 &
bursts / invented endpoints \\

\addlinespace

WAF attack flags &
0.5455 & 0.1818 & 0.6897 & 0.7273 & 0.0968 &
flagged burst poison \\

\addlinespace

HTTP status code &
0.5455 & 0.1818 & 0.7586 & 0.7273 & 0.0968 &
unsuccessful / flagged poison \\

\addlinespace

Value frequency floor &
0.2727 & 0.2424 & 1.0000 & 0.7433 & 0.1494 &
stealthy injection in established fields \\

\addlinespace

All seven candidate signals &
0.1111 & 0.0909 & 0.3448 & 0.9000 & n/a &
combined coverage \\

\bottomrule
\end{tabularx}
\end{table*}

Removing the value frequency floor from the configuration with all seven signals is the only single removal that reduces effectiveness, from 0.9000 to 0.7740. The value frequency floor combined with either WAF attack flags or HTTP status reaches effectiveness 0.9091 on the balanced profile. The seven signal configuration is retained for the ablation analysis because it maintains effectiveness of 0.9000 across all five tested attacker profiles.

Representative pair interactions are reported in Table~\ref{tab:pairs}. The per IP cap contributes zero in these experiments and remains disabled in the default WAPP configuration, which therefore has six active signals. A sensitivity analysis gives the same result for value frequency thresholds of 0.70, 0.80, 0.90, and 0.95, as reported in Table~\ref{tab:floorlattice}. This indicates that the measured benefit comes mainly from enabling the floor rather than from the exact cutoff.

\begin{strip}
\centering
\captionof{table}{RQ2 representative signal pairs}
\label{tab:pairs}
\small
\setlength{\tabcolsep}{6pt}

\begin{tabularx}{\textwidth}{
@{\hspace{0.03\textwidth}}
>{\raggedright\arraybackslash}p{0.28\textwidth}
@{\hspace{0.06\textwidth}}
>{\centering\arraybackslash}p{0.14\textwidth}
@{\hspace{0.14\textwidth}}
>{\raggedright\arraybackslash}p{0.39\textwidth}}
\toprule
\textbf{Signal pair} &
\textbf{Effectiveness} &
\textbf{Note} \\
\midrule

Floor + WAF attack flags &
0.9091 &
best pair; slightly exceeds the seven signal result \\

Floor + HTTP status code &
0.9091 &
tied best pair \\

Floor + temporal spread &
0.7193 &
limited additional benefit from temporal spread \\

WAF flags + HTTP status &
0.7273 &
two flag based signals are redundant on this dataset \\

Distinct IP + observation count &
0.5692 &
both gates affect the same low evidence endpoints \\

Per IP cap + any single gate &
$=$ that gate &
cap contributes zero \\

\bottomrule
\end{tabularx}
\end{strip}

\begin{table*}[!ht]
\centering
\caption{RQ2 value frequency floor sensitivity}
\label{tab:floorlattice}
\small

\begin{tabularx}{\textwidth}{@{\hspace{0.03\textwidth}}lYYYY}
\toprule
\textbf{Floor} &
\textbf{WCR} &
\textbf{PER} &
\textbf{Legit FPR} &
\textbf{Effectiveness} \\
\midrule

Off  & 0.4444 & 0.1515 & 0.0222 & 0.7740 \\
0.70 & 0.1111 & 0.0909 & 0.0222 & 0.9000 \\
0.80 & 0.1111 & 0.0909 & 0.0222 & 0.9000 \\
0.90 (default) & 0.1111 & 0.0909 & 0.0222 & 0.9000 \\
0.95 & 0.1111 & 0.0909 & 0.0222 & 0.9000 \\

\bottomrule
\end{tabularx}
\end{table*}

RQ2 also evaluates an adaptive attacker designed to satisfy the source diversity and temporal gates. Poison is distributed across 60 source IPs and sustained beyond the 12 hour requirement while attempting to introduce a target metacharacter into the learned class. Across the 20 tested cells, the targeted metacharacter frequency reaches at most 0.50, remaining below the 0.90 value frequency floor. Maximum PER and WCR are therefore 0, while the benign control admit rate remains 1.0. This result applies only to the tested gate constraints and does not imply resistance to all adaptive poisoning strategies.

Across five attacker profiles, the configuration with all seven candidate signals maintains effectiveness of 0.9000. The empty filter ranges from 0.4545 to 0.6791, while the Kruegel--Vigna baseline ranges from 0.5758 to 0.7107, as summarized in Table~\ref{tab:rq2-adaptive-results}.

\begin{table}[!ht]
\centering
\caption{RQ2 adaptive attack results}
\label{tab:rq2-adaptive-results}
\small
\setlength{\tabcolsep}{3pt}

\begin{tabularx}{\columnwidth}{
>{\raggedright\arraybackslash}p{105pt}
>{\raggedright\arraybackslash}X}
\toprule
\textbf{Measure} &
\textbf{RQ2 result} \\
\midrule

Adaptive cells &
20 \\

\addlinespace

Topology constraints &
60 distinct IPs (gate $\geq 50$) and temporal spread $\geq 12$ hours \\

\addlinespace

Maximum targeted metacharacter frequency &
0.50, below the 0.90 value frequency floor \\

\addlinespace

Maximum PER / WCR &
0.00 / 0.00 \\

\addlinespace

Benign control admit rate &
1.00 \\

\addlinespace

All seven signal effectiveness &
0.9000 across five attacker profiles \\

\addlinespace

Baselines &
Empty filter: 0.4545--0.6791; Kruegel--Vigna: 0.5758--0.7107 \\

\bottomrule
\end{tabularx}
\end{table}

\subsection{Statistical Synthesis Is Selected as the Default Synthesis Engine}

RQ3 compares the deterministic statistical synthesizer with a fixed qwen/qwen3.5-27b LLM configuration. Both receive identical trusted profiles and use the same rule schema, conversion, live WAF replay, ten attack payloads, and disjoint legitimate holdouts. The comparison therefore isolates the synthesis method.

The endpoint results in Table~\ref{tab:static_vs_llm} show a tradeoff between security and precision. The LLM reduces false positives on the DVWA username, Juice Shop search, and Airport comment fields. However, on the Airport comment field it admits two attack payloads (PER 0.20), while the statistical engine blocks all ten attacks but produces a higher FPR. Both approaches perform identically on the Juice Shop email field and Airport item lookup. At the application level, statistical synthesis achieves PER/FPR of 0.00/0.76 on DVWA, 0.00/0.07 on Juice Shop, and 0.10/0.425 on Airport, compared with 0.00/0.00, 0.00/0.00, and 0.20/0.13 for the LLM.

The engine choice is based on both measured performance and structural properties. The paired attack blocking difference between the statistical and LLM approaches is +0.057 with a bootstrap 95\% CI of [0.0, 0.114], which does not show a clear attack blocking advantage. For legitimate traffic, the difference is +0.297 with a 95\% CI of [0.10, 0.51], showing lower FPR for the LLM. No ungrounded constraints were observed across the 15 LLM runs, while the statistical engine is grounded by construction. Closed parameter set enforcement appears in 6 of 7 statistical rules and in none of the 15 LLM runs. The statistical engine has zero model inference cost, compared with \$0.2747 for the 15 LLM runs, whose mean generation times are 66.9 s for DVWA, 98.9 s for Juice Shop, and 197.2 s for Airport.

The two Airport comment endpoints represent the same comment field in JSON and form formats and produce identical results. They are therefore not fully independent observations. The bootstrap intervals are reported descriptively as effect size context rather than as an inferential test over independent endpoints \cite{efron1979bootstrap}.

\begin{table*}[!t]
\centering
\caption{RQ3 statistical and LLM comparison}
\label{tab:static_vs_llm}
\small
\setlength{\tabcolsep}{1.8pt}

\begin{tabularx}{\textwidth}{
@{\hspace{0.03\textwidth}}
>{\raggedright\arraybackslash}p{0.25\textwidth}
>{\centering\arraybackslash}X
>{\centering\arraybackslash}X
>{\centering\arraybackslash}X
>{\centering\arraybackslash}X
@{}
}
\toprule
\textbf{Endpoint / field} &
\shortstack{\textbf{Static}\\\textbf{PER}} &
\shortstack{\textbf{Static}\\\textbf{FPR}} &
\shortstack{\textbf{LLM}\\\textbf{PER}} &
\shortstack{\textbf{LLM}\\\textbf{FPR}} \\
\midrule

DVWA login / username &
0.00 & 0.76 (38/50) & 0.00 & 0.00 (0/50) \\

Juice Shop search / q &
0.00 & 0.14 (7/50) & 0.00 & 0.00 (0/50) \\

Juice Shop login / email &
0.00 & 0.00 (0/50) & 0.00 & 0.00 (0/50) \\

Airport comment / JSON &
0.00 & 0.78 (39/50) & 0.20 & 0.26 (13/50) \\

Airport comment / form &
0.00 & 0.78 (39/50) & 0.20 & 0.26 (13/50) \\

Airport search / q &
0.00 & 0.14 (7/50) & 0.00 & 0.00 (0/50) \\

Airport item lookup / id &
0.40 (4/10 admit) & 0.00 (0/50) &
0.40 (4/10 admit) & 0.00 (0/50) \\

\bottomrule
\end{tabularx}
\end{table*}

Based on these results, the statistical engine is selected as the default synthesizer because it provides grounded, closed set rules with zero model inference cost and no observed attack blocking disadvantage. Higher false positives on punctuation rich fields remain a known limitation. The LLM may be used as an operator aid for suggesting field level character sets, subject to human approval.

A wider validation tests the locked statistical engine across nine DVWA modules and ten Juice Shop endpoints through live reverse proxy replay. Solved Juice Shop captchas provide genuine WAF verdicts. The application level PER and FPR results are summarized in Figure~\ref{fig:app_performance}.

\begin{center}
\begin{minipage}{0.95\columnwidth}
    \centering
    \includegraphics[width=\linewidth]{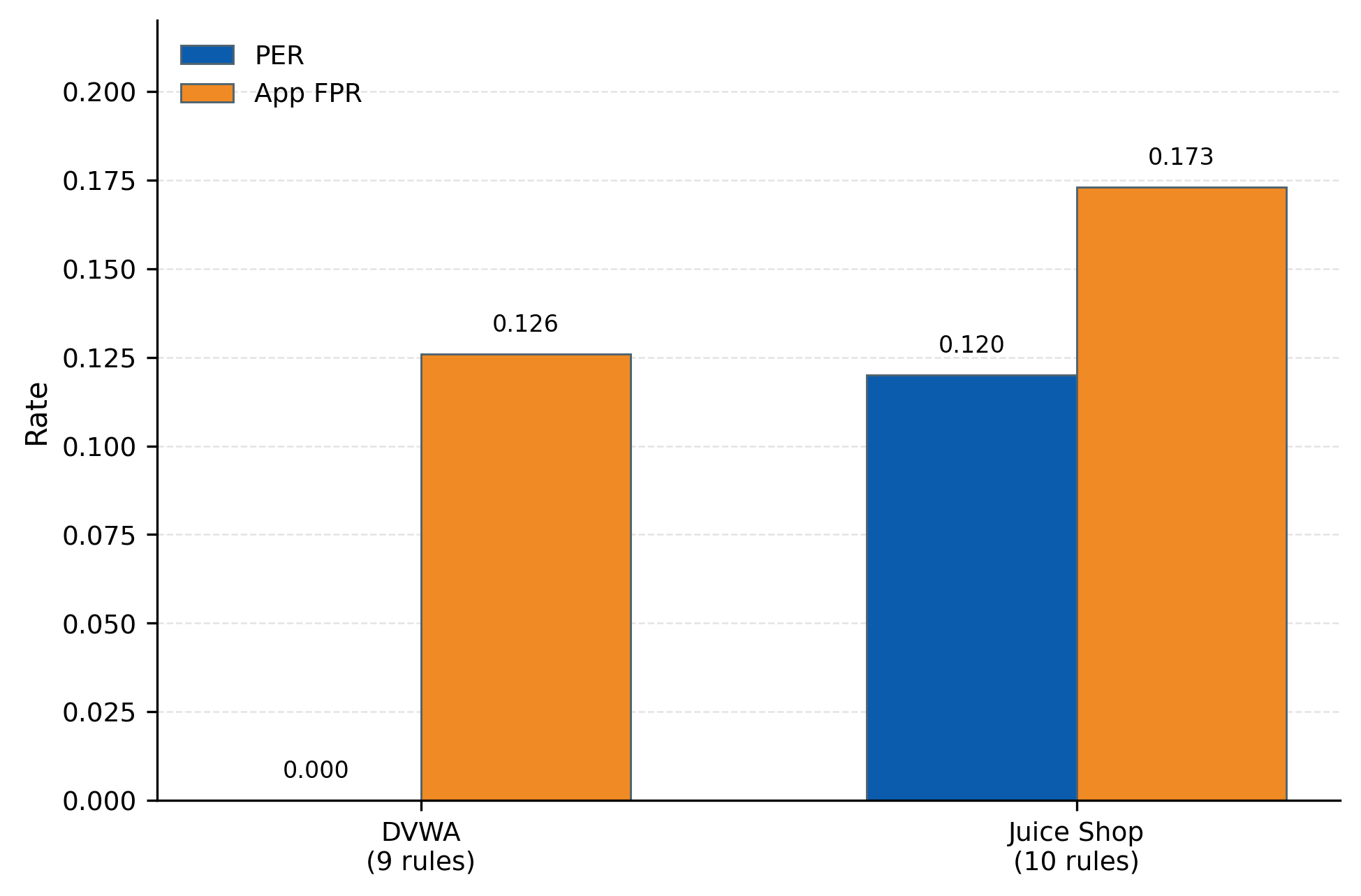}
    \captionof{figure}{RQ3 full surface validation}
    \label{fig:app_performance}
\end{minipage}
\end{center}

DVWA achieves PER 0.00 with an application wide FPR of 0.126, while Juice Shop achieves PER 0.12 (12/100) with an FPR of 0.173. All 12 Juice Shop attack admissions occur across four integer ID path endpoints, with three admissions per endpoint, when the payloads leave the learned path segments and no longer match the endpoint rule. This behavior results from per endpoint path matching rather than acceptance by the integer constraint itself.

False positives are concentrated in punctuation rich text fields. Constrained and integer ID fields maintain FPR 0.00, while text field FPR ranges from 0.33 to 0.47. These results indicate that the precision issue is concentrated in rich text fields rather than across all endpoint types, although the finding remains limited to the two evaluated applications.

\subsection{Evidence Scores Require Representative Validation}

RQ4 evaluates whether the evidence score relates to false positive behavior during live WAF rehearsal. On the controlled synthetic holdout, all rules scoring 0.4787 or below blocked 60 of 108 legitimate values, while all rules scoring 0.5873 or above produced no false positives, as shown in Table~\ref{tab:score_tier_evaluation}. This observed separation suggests a relationship between evidence score and false positive behavior on this controlled gradient.

An exploratory Fisher's exact test on the endpoint level false positive outcome, comparing scores below and at or above 0.5873, gives $p=0.004545$ \cite{fisher_interpretation_1922}. Because the 0.5873 cutoff is identified from the same calibration data, this test is treated as exploratory rather than as independent validation of the cutoff. The value 0.5873 is therefore used as an advisory calibration cutoff, while representative validation remains necessary before enforcement.

\begin{table}[!htt]
\centering
\caption{RQ4 confidence score calibration}
\label{tab:score_tier_evaluation}
\small
\setlength{\tabcolsep}{2pt}
\begin{tabularx}{\linewidth}{
c
c
>{\centering\arraybackslash}X
>{\centering\arraybackslash}X
>{\centering\arraybackslash}X
}
\toprule
\textbf{Score} &
\textbf{Tier} &
\textbf{False positives / 108} &
\textbf{Genuine attack block} &
\textbf{Cells / guardrail} \\
\midrule
0.0000 & floor  & 60 (55.6\%) & 100\% & 1 / violated \\
0.4287 & low    & 60 (55.6\%) & 100\% & 1 / violated \\
0.4787 & low    & 60 (55.6\%) & 100\% & 1 / violated \\
0.5873 & medium & 0 (0\%)     & 100\% & 3 / met \\
0.7709 & high   & 0 (0\%)     & 100\% & 3 / met \\
0.8271 & rich   & 0 (0\%)     & 100\% & 3 / met \\
\bottomrule
\end{tabularx}
\end{table}
\FloatBarrier

A representative DVWA username holdout demonstrates the limitation of relying on evidence score alone. As shown in Table~\ref{tab:vf_floor_sensitivity}, the rule retains a high pre-penalty evidence score of 0.7709, but at the default 0.90 value frequency floor it blocks 28 of 43 legitimate usernames (FPR 0.6512), despite blocking all 13 genuine attack vectors.

\begin{center}
\begin{minipage}{\columnwidth}
\centering
\captionof{table}{RQ4 DVWA value frequency floor sensitivity}
\label{tab:vf_floor_sensitivity}
\scriptsize
\setlength{\tabcolsep}{1.5pt}
\renewcommand{\arraystretch}{1.15}

\begin{tabularx}{\columnwidth}{
c
>{\centering\arraybackslash}X
>{\centering\arraybackslash}X
>{\centering\arraybackslash}X
c
}
\toprule
\shortstack{\textbf{Value}\\\textbf{freq. floor}} &
\shortstack{\textbf{FP / 43}\\\textbf{(FPR)}} &
\shortstack{\textbf{Genuine}\\\textbf{attack block}} &
\shortstack{\textbf{17 item}\\\textbf{evasion}} &
\textbf{Guardrail} \\
\midrule

0.90 & 28/43 (0.6512) & 100\% & 0.0000 & Violated \\
0.70 & 28/43 (0.6512) & 100\% & 0.0000 & Violated \\
0.45 & 28/43 (0.6512) & 100\% & 0.0000 & Violated \\
0.20 & 13/43 (0.3023) & 100\% & 0.0588 & Violated \\
0.10 & 0/43 (0.0000)  & 100\% & 0.1176 & Met \\

\bottomrule
\end{tabularx}
\end{minipage}
\end{center}

Lowering the floor to 0.10 reduces FPR to 0 while still blocking all 13 genuine attacks. However, two non exploit boundary probes from the wider 17 item rehearsal are admitted, giving a 17 item suite evasion rate of 0.1176. Validation feedback then reduces the rule score to 0.5709, below the 0.5873 advisory cutoff.

All generated whitelist rules compiled and loaded successfully on the three routed targets (10, 1, and 1 rules). The reported 100\% attack block rate refers to the 13 genuine attack vectors. The wider 17 item rehearsal additionally includes four non exploit boundary and mimicry probes, which are evaluated separately when applying the validation feedback penalty.

\subsection{Learned Positive Security Blocks Confirmed Signature Bypasses on Constrained Fields}

RQ5 first establishes the signature baseline. At the local anomaly threshold of 100, all five canonical attacks are admitted, while at the CRS recommended threshold of 5, all five are blocked. Eleven same class bypass candidates are then tested at threshold 5. Five bypass CRS: all three XSS variants and two of three path traversal variants, while none of the five SQL injection variants evade detection. These five HTTP 200 responses form the confirmed bypass set, as summarized in Table~\ref{tab:attack_class_thresholds} \cite{portswigger_web_security_academy_cross-site_nodate,portswigger_web_security_academy_file_nodate}.

\begin{center}
\begin{minipage}{\columnwidth}
\centering
\captionof{table}{RQ5 signature and bypass results}
\label{tab:attack_class_thresholds}
\scriptsize
\setlength{\tabcolsep}{2pt}
\renewcommand{\arraystretch}{1.12}

\begin{tabularx}{\columnwidth}{
>{\raggedright\arraybackslash}p{58pt}
>{\centering\arraybackslash}X
>{\centering\arraybackslash}X
>{\centering\arraybackslash}X
}
\toprule
\textbf{Attack class} &
\shortstack{\textbf{Canonical}\\\textbf{threshold 100}} &
\shortstack{\textbf{Canonical}\\\textbf{threshold 5}} &
\shortstack{\textbf{Bypass evasion}\\\textbf{threshold 5}} \\
\midrule

SQL injection &
2/2 admitted (200) &
2/2 blocked (403) &
0/5 evaded \\

Cross site scripting &
2/2 admitted (200) &
2/2 blocked (403) &
3/3 evaded \\

Path traversal &
1/1 admitted (200) &
1/1 blocked (403) &
2/3 evaded \\

\bottomrule
\end{tabularx}
\end{minipage}
\end{center}

The five confirmed bypasses are then replayed against two constrained Airport fields: an alphanumeric plus space search term and a signed integer page selector. Each bypass is tested once on each field, producing ten attacker requests. At threshold 100, where the signature layer admits these bypasses, the learned constraints block all 10 requests with HTTP 403. The same 10 requests are also blocked in the layered threshold 5 configuration. Both in shape controls (2/2), ``apple juice'' and page value ``2'', remain admitted with HTTP 200.

Table~\ref{tab:out_of_shape_analysis} shows that the additional blocking results from field shape enforcement rather than attack name recognition or route level blocking.

\begin{strip}
\centering
\captionof{table}{RQ5 blocking of confirmed CRS bypasses}
\label{tab:out_of_shape_analysis}
\footnotesize
\setlength{\tabcolsep}{3pt}
\renewcommand{\arraystretch}{1.08}

\begin{tabularx}{\textwidth}{
>{\raggedright\arraybackslash}p{0.27\textwidth}
>{\raggedright\arraybackslash}p{0.25\textwidth}
>{\centering\arraybackslash}X
>{\centering\arraybackslash}X
}
\toprule
\textbf{Payload / field} &
\textbf{Out of shape character(s)} &
\textbf{Whitelist only arm (threshold 100)} &
\textbf{Layered arm (threshold 5)} \\
\midrule

Attribute breakout / search &
double quote &
403 blocked &
403 blocked \\

JavaScript context breakout / search &
apostrophe, parenthesis, semicolon &
403 blocked &
403 blocked \\

Template literal breakout / search &
backtick, parenthesis &
403 blocked &
403 blocked \\

Doubled slash traversal / search &
period, slash &
403 blocked &
403 blocked \\

Doubled backslash traversal / search &
period, backslash &
403 blocked &
403 blocked \\

\addlinespace

Attribute breakout / page &
non digit characters &
403 blocked &
403 blocked \\

JavaScript context breakout / page &
non digit characters &
403 blocked &
403 blocked \\

Template literal breakout / page &
non digit characters &
403 blocked &
403 blocked \\

Doubled slash traversal / page &
non digit characters &
403 blocked &
403 blocked \\

Doubled backslash traversal / page &
non digit characters &
403 blocked &
403 blocked \\

\addlinespace

Negative control: ``apple juice'' / search &
none; in shape &
200 admitted &
200 admitted \\

Negative control: ``2'' / page &
none; in shape &
200 admitted &
200 admitted \\

\bottomrule
\end{tabularx}
\end{strip}

Free text fields show the corresponding precision limitation. On the Airport comment corpus, the 0.90 value frequency floor admits only special characters appearing in at least 90\% of legitimate values. The space appears in 100\% of values and is admitted, while the period (73\%), comma (47\%), apostrophe (27\%), exclamation mark (20\%), hyphen (13\%), and slash (7\%) remain excluded.

The resulting rule blocks all four attacks tested on this field, but it also blocks the benign comment ``Great product, arrived on time!'' because its comma and exclamation mark remain outside the learned shape. This experiment demonstrates the precision limitation but does not estimate a general free text FPR from a full benign holdout.

As shown in Table~\ref{tab:character_analysis}, enabling only the period changes the legitimate comment ``Great product. Thanks'' from HTTP 403 to 201, while the traversal payload remains blocked because the slash is still excluded. This demonstrates field specific character adjustment as a precision control, although each additional character must be evaluated separately \cite{owasp_cheat_sheet_series_input_nodate,owasp_cheat_sheet_series_virtual_nodate}.

\begin{center}
\begin{minipage}{\columnwidth}
\centering
\captionof{table}{RQ5 free text character analysis}
\label{tab:character_analysis}
\scriptsize
\setlength{\tabcolsep}{1.5pt}
\renewcommand{\arraystretch}{1.08}

\begin{tabularx}{\columnwidth}{
>{\raggedright\arraybackslash}p{43pt}
>{\centering\arraybackslash}X
>{\centering\arraybackslash}p{38pt}
>{\centering\arraybackslash}p{40pt}
>{\raggedright\arraybackslash}X
}
\toprule
\textbf{Character} &
\shortstack{\textbf{Legitimate}\\\textbf{frequency}} &
\shortstack{\textbf{$\geq$ 0.90}\\\textbf{floor}} &
\shortstack{\textbf{Learned}\\\textbf{shape}} &
\textbf{Action} \\
\midrule

space &
100\% &
Yes &
Admitted &
None \\

period &
73\% &
No &
Excluded &
Enabled; 403 $\rightarrow$ 201 \\

comma &
47\% &
No &
Excluded &
None \\

apostrophe &
27\% &
No &
Excluded &
None \\

exclamation mark &
20\% &
No &
Excluded &
None \\

hyphen &
13\% &
No &
Excluded &
None \\

slash &
7\% &
No &
Excluded &
Kept excluded; traversal 403 \\

\bottomrule
\end{tabularx}
\end{minipage}
\end{center}

\section{Discussion}
\label{sec:discussion}

The five research questions show that learned positive security depends on two main conditions: the training evidence must be trustworthy, and the protected field must have a shape that can be constrained without rejecting legitimate use. RQ1 demonstrates the risk of learning from untrusted traffic, while RQ3--RQ5 show the limitations of applying strict constraints to flexible fields. WAPP therefore separates learning from enforcement through trust filtering, representative validation, and operator approval.

The confidence score should be interpreted as evidence support rather than a guarantee of rule safety. Rules below the advisory cutoff should remain under monitoring or review. Constrained fields with sufficient evidence and successful representative validation are better candidates for enforcement, while free text fields require broader legitimate holdouts and field specific character controls. The multi signal trust design is retained because the configuration with all seven candidate signals maintains effectiveness of 0.9000 across all five tested attacker profiles. The per IP cap remains disabled in the default configuration because it provides no measured improvement. A smaller two signal combination reaches slightly higher effectiveness on the balanced profile, but this result does not extend across the full attacker set.

The LLM comparison leads to a similarly limited conclusion. The tested LLM preserves legitimate punctuation better, but provides no clear attack blocking advantage and does not generate the closed parameter set constraints observed in 6 of 7 statistical rules. The statistical engine is therefore selected as the default synthesizer, while the LLM is better suited to operator advice. This conclusion applies only to the tested LLM configuration and evaluated fields and should not be generalized to all language model based policy generation.

\section{Conclusion and Future Work}
\label{sec:conclusion}

WAPP shows that learning positive security policies from live traffic requires trusted training data and validation before enforcement. On the tested DVWA username field, unfiltered learning becomes Degraded at 0.2\% poisoned traffic and Broken at 0.5\%, while the evaluated free text field admits malicious inputs even without poisoning. On the frozen RQ2 dataset, the configuration with all seven candidate signals increases measured poisoning resilience effectiveness from 52.88\% with no filtering to 90.00\%, compared with 62.31\% for the Kruegel--Vigna baseline.

The statistical synthesizer is selected because it is deterministic, has no model inference cost, and produces closed parameter set enforcement in 6 of 7 evaluated rules without an observed attack blocking disadvantage compared with the tested LLM. Confidence scoring provides a useful measure of evidence support, but representative validation remains necessary to identify false positives before enforcement. Finally, 45.45\% (5 of 11) of the tested CRS bypass candidates are confirmed, and the learned constraints block all 10 resulting bypass requests across the two constrained fields while admitting both in shape controls (2/2).

The results are limited to the tested applications, traffic, WAF configuration, attacker profiles, and one LLM setup. Future work should evaluate WAPP across more applications, independent seeds, production like traffic, additional WAF and CRS configurations, and multiple language models. Further work should also improve concept drift handling, free text policies, and confidence measures for field shape coverage.

\begin{acknowledgement}
This work was supported by the security research and development department at Cyshield Company, Cairo, Egypt.
\end{acknowledgement}

\section*{Statements and Declarations}

\noindent\textbf{Competing interests.} The authors declare that they have no competing interests.

\noindent\textbf{Data availability.} The experimental data and code supporting the findings of this study are available from the authors upon reasonable request.

\bibliographystyle{spphys}
\bibliography{WAPP_References}

\end{document}